\documentclass[11pt]{article}
\usepackage[left=3.2cm,top=3.2cm,bottom=3.2cm,right=3.2cm,headsep=5pt, footskip=20pt]{geometry}

\usepackage[round,comma]{natbib}
\usepackage{xcolor}
\usepackage{graphicx}
\graphicspath{ {./} }

\usepackage{hyperref}
\hypersetup{
    colorlinks=false,
    linkcolor=black,
    filecolor=black,      
    urlcolor=black,
}
 
\makeatletter
\newcommand{\printfnsymbol}[1]{%
  \textsuperscript{\@fnsymbol{#1}}%
}
\makeatother

\usepackage{authblk}

\begin{document}
\title{Analyzing biological and artificial neural networks: challenges with opportunities for synergy?}

\author{\textbf{David G.T. Barrett}\thanks{equal contribution, alphabetical order}$~^{1}$, \textbf{Ari S. Morcos}\printfnsymbol{1}\thanks{current address for A.S. Morcos: Facebook AI Research (FAIR), Menlo Park, CA, USA}$~^{1}$
 and \textbf{Jakob H. Macke}$^{2}$ }
\date{%
    $^1$\small DeepMind, London, UK\\%
    $^2$Computational Neuroengineering, Department of Electrical and Computer Engineering,\\ Technical University of Munich, Germany\normalsize \newline  \\%
    \today
}

\maketitle

\begin{abstract}

Deep neural networks (DNNs) transform stimuli across multiple processing stages to produce representations that can be used to solve complex tasks, such as object recognition in images. However, a full understanding of how they achieve this remains elusive. The complexity of biological neural networks substantially exceeds the complexity of DNNs, making it even more challenging to understand the representations that they learn. Thus, both machine learning and computational neuroscience are faced with a shared challenge: how can we analyze their representations in order to understand how they solve complex tasks?

We review how data-analysis concepts and techniques  developed by computational neuroscientists can be useful for analyzing representations in DNNs, and in turn, how recently developed techniques for analysis of DNNs can be useful for understanding representations in biological neural networks. We explore opportunities for synergy between the two fields, such as the use of DNNs as in-silico model systems for neuroscience, and how this synergy can lead to new hypotheses about the operating principles of biological neural networks.
\end{abstract}



\section{Introduction}

Neuroscience is in the midst of a technological transformation, enabling us to investigate the structure and function of neural circuits at unprecedented scale and resolution. Electrophysiological technologies \citep{jun2017fully} and imaging tools \citep{ahrens2012brain} have made it possible to record the activity of hundreds of neurons simultaneously, and opto-genetic techniques enable targeted perturbations of neural activity \citep{packer2014simultaneous,lerman2018spatially}. These advances open up the possibility of empirical investigation into the computations that are distributed across large neural populations and hold the promise of providing fundamental insights into how populations of neurons collectively perform computations. However, it has also become increasingly clear that interpreting the complex data generated by these modern experimental techniques, and distilling a deeper understanding of neural computation is a challenging problem which requires powerful analysis tools \citep{cunningham2014dimensionality}.

In parallel, the field of machine learning is undergoing a transformation, driven by advances in deep learning. This has lead to a large increase in the performance and widespread use of DNNs across numerous diverse problem domains such as object recognition \citep{krizhevsky2012imagenet,simonyan2014very}, automated language translation \citep{wu2016google}, game-play \citep{mnih2015human,silver2017mastering} and scientific applications \citep{januszewski2018high-precision}. Deep networks consist of large numbers of  linearly connected nonlinear units whose parameters are tuned using numerical optimization. Neuroscience and cognitive science were influential in the early development of DNNs \citep{rumelhart1987parallel} and convolutional neural networks (CNNs), widely used in computer vision \citep{fukushima1982neocognitron,lecun1990handwritten,krizhevsky2012imagenet}, were inspired by canonical properties of the ventral visual stream. 

Even though we have full access to DNNs which allows us to measure complete connectivity and complete activation patterns, it has nonetheless been challenging to develop a theoretical understanding of how and why they work.  A deeper understanding of DNNs is important for application domains where interpretable DNNs are required and is also important for guiding the development of better model architectures. One reason that it is difficult to understand DNNs  is that they usually contain millions of parameters. For example, `AlexNet', which is well known for having demonstrated the potential of CNNs, contains 8 layers and a total of 60 million parameters \citep{krizhevsky2012imagenet}. Modern state of the art networks are often much larger. We still do not fully understand how and why DNNs can generalize so well without overfitting \citep{zhang2017understanding,dinh2017sharp}, nor do we fully understand how invariant representations arise in these multi-layer networks
\citep{tishby2015deep,achille2017emergence}. 

Therefore, both neuroscience and deep learning face a similar challenge: how do neural networks, consisting of large numbers of interconnected elements, transform representations of stimuli across multiple processing stages so as to implement a wide range of complex computations and behaviours, such as object recognition? What data-analysis techniques are most useful in this endeavor? How can we characterize and analyze representations in high-dimensional spaces? These common challenges open up opportunities for synergy in analysis across neuroscience and machine learning \citep{victor2005analyzing}. In this review, we explore statistical tools from both of these disciplines that provide insight into neural representation and computation by analyzing activity-measurements of both single neurons and neural populations.

\section{Receptive fields: What stimulus-features do single neurons represent?}

\begin{figure}
    \centering
    \includegraphics[width=0.99\textwidth]{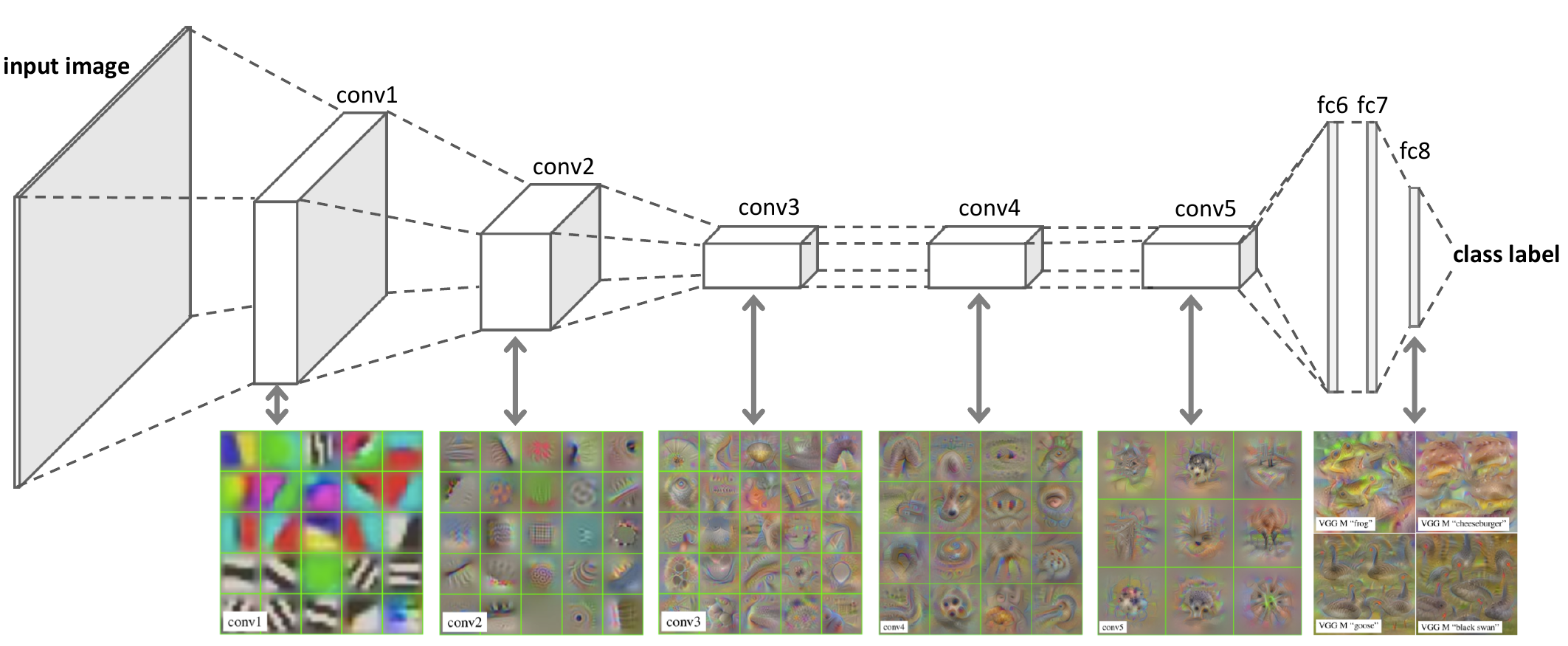}
    \caption{{\bf Receptive field analysis of neural filters from a DNN known as VGG-M.} A schematic of the VGG-M network (top) illustrates how an input image is progressively processed through banks of convolutional (conv) filters, followed by a series of fully connected (fc) layers before finally producing a class label as output. Receptive fields (bottom) for each layer  are calculated using activation maximisation \citep{mahendran2016visualizing}. The receptive fields become increasingly complex, with earlier units having edge-like structure, intermediate layers responding to complex textures and the final layer responding to object-like images. It is not always possible to find a semantically meaningful description of a receptive field, as can be seen for the intermediate layers. \emph{Adapted by permission from Springer Nature: Springer 	International Journal of Computer Vision Visualizing, Deep Convolutional Neural Networks Using Natural Pre-images, Aravindh Mahendran and Andrea Vedaldi, 2016}.}
    \label{fig:ReceptiveFields}
\end{figure}

Receptive field analysis is the canonical method in neuroscience for analyzing single neuron activity \citep{Sherrington1906, victor2005analyzing}. The receptive field of a neuron usually refers to the region of stimulus space that causes a neural firing response. More generally, it is often characterized as the `preferred stimulus' for a neuron - the stimulus that elicits a maximal response. Perhaps the most famous receptive field analysis was the Nobel prize winning work of Hubel and Wiesel, who discovered that the receptive fields of simple cells in the visual cortex have a spatially localized edge-like structure \citep{hubel1959receptive}. Our canonical understanding of visual processing has been largely informed by receptive field analysis. According to this perspective, the receptive fields of neurons along the ventral stream of the visual processing pathway become progressively larger \citep{yamins2016using}, increasingly complex  \citep{pasupathy2001shape,gucclu2015deep}
and increasingly invariant to changes in the input-statistics  \citep{rust2010selectivity}, culminating in concept cells that are tuned to individual objects, but largely invariant to their visual appearance \citep{Quiroga2005}.

In recent years, receptive field analysis has also become a canonical method for analyzing response properties in artificial neural networks \citep{zeiler2014visualizing,yosinski2015understanding,mahendran2015understanding,luo2016understanding,mahendran2016visualizing}. For example, a receptive field analysis method known as activation maximization \citep{mahendran2016visualizing,nguyen2016synthesizing,cadena2018diverse} can be used to synthesize images which maximally activate units in artificial neural networks (Fig. \ref{fig:ReceptiveFields}) \citep{mahendran2016visualizing}. Similar to the ventral pathway, the receptive fields of units in CNNs become progressively larger \citep{luo2016understanding} and increasingly complex. Earlier units having edge-like receptive fields and later units respond to more complex textures \citep{zeiler2014visualizing,mahendran2016visualizing}. The units of the final layer have class-specific receptive fields (Fig. \ref{fig:ReceptiveFields}) which correspond to specific object categories, akin to concept-cells \citep{Quiroga2005,Quoc2012,mahendran2016visualizing}.

\section{Ablations: How relevant are single neurons to the overall computation?}

\begin{figure}
    \centering
    \includegraphics[width=0.99\textwidth]{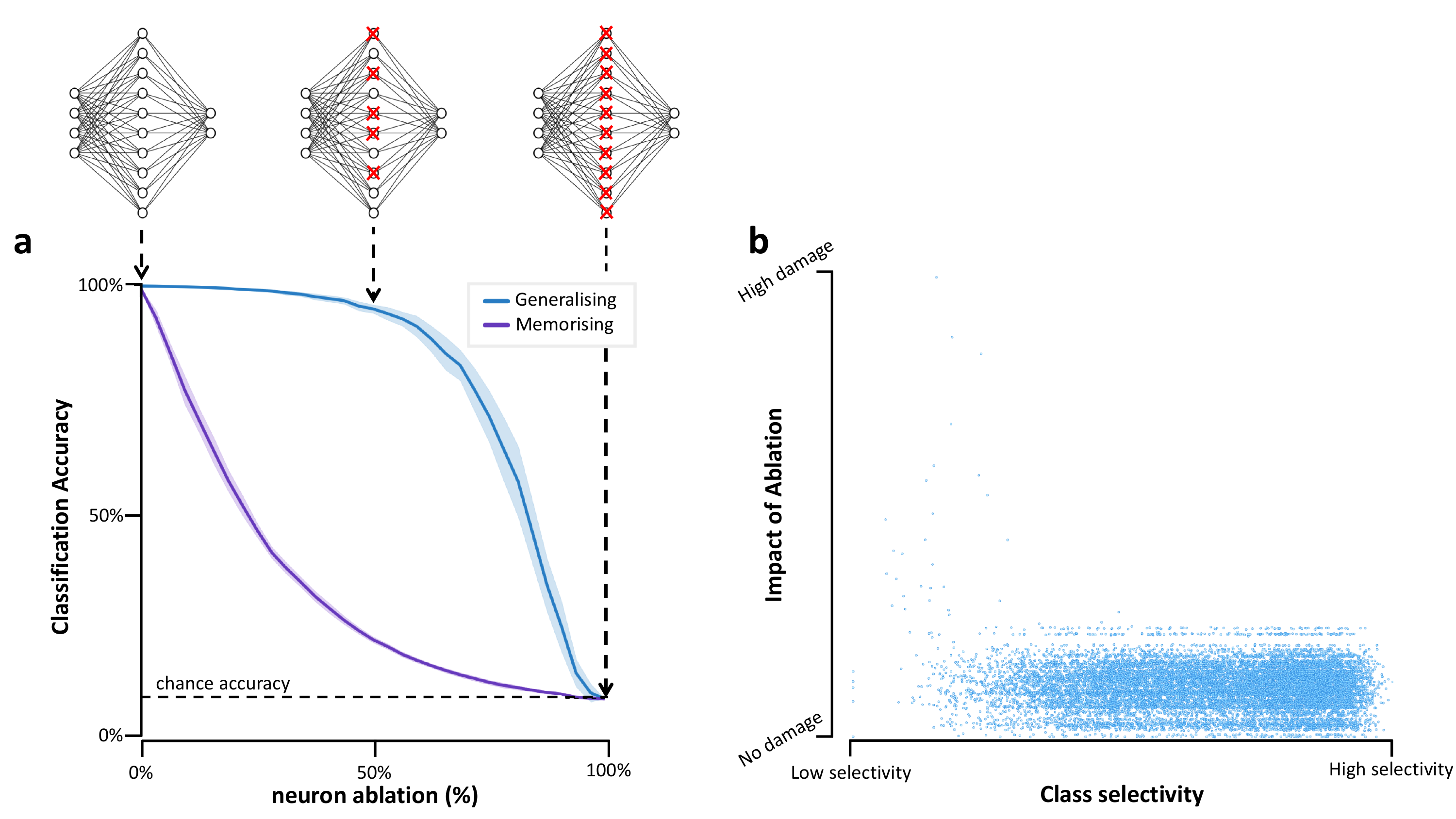}
        \caption{\label{fig:ablation}
        {\bf Impact of neural ablations on DNN accuracy}
        {\bf a)} By ablating neurons and measuring DNN classification accuracy, the impact of neural ablations on DNN performance can be evaluated. A network with good generalization ability (cyan) is much more robust to ablation than networks which overfit the data using memorisation strategies (purple). {\bf b)} The relationship between the class selectivity of a neuron and the importance of a neuron for classification can be evaluated by measuring the impact of neuron ablation on classification accuracy as a function of the neuron class selectivity index. Most neurons have little impact on accuracy, and neuron class selectivity does not correlate strongly with the impact of neuron ablation. Figures reproduced with modification from \citep{morcos2018importance}.
        }
\end{figure}

Historically, neuroscience has focused on single neurons with clearly defined tuning properties, in part, due to the technical constraints involved in recording neural activity. However, a number of recent studies have demonstrated that neurons with clearly defined tuning, such as visual cortex simple cells, are relatively uncommon and that neurons with ambiguous tuning properties often contain substantial amounts of task-relevant information, especially when analyzed as populations \citep{Olshausen2006,rigotti2013importance, ManteSussillo_13, raposo2014category, goris2015, morcos2016history}. This leads to the question: how important are selective, clearly tuned neurons?

Ablation analysis allows us to answer this question by silencing neurons and measuring the impact of this intervention on network output. It is difficult to perform single-neuron ablation analysis in biological neural networks, due to technical constraints \citep{Theunissen1991}, although this may change in the coming years due to the advent of optogenetics and targeted two-photon stimulation \citep{packer2014simultaneous,lerman2018spatially}. In theoretical neuroscience single-neuron ablation analysis is possible, and has allowed us to understand the impact of ablations on neural computation in model systems of both intact and damaged biological networks \citep{Barrett2016}.

Ablation analysis is also possible in DNNs, as we can perfectly characterize the activity of every neuron in response to any arbitrary ablation. For example, ablation was recently used to evaluate the relationship between unit selectivity and unit importance as well as the relationship between network robustness and generalization in CNNs performing image classification tasks \citep{morcos2018importance}. By measuring how network accuracy dropped as increasing numbers of neurons were deleted, it was found that networks which learn generalizable solutions (i.e. those solutions which generalize to images never seen during training) were more robust to ablations than those which simply memorize the training data (Fig. \ref{fig:ablation} a). 
Leveraging the ability to simultaneously characterize unit selectivity and unit importance, the impact of ablation on network accuracy and on the class selectivity of each unit was calculated. Perhaps surprisingly, there was little relationship between class selectivity and importance, suggesting that units critical to network computation were just as likely to exhibit ambiguous tuning as they were to be clearly tuned (Fig. \ref{fig:ablation} b). If the maximum drop of accuracy for \emph{any} class is calculated rather than the drop in overall accuracy \citep{zhou2018revisiting}, a significant but relatively weak ($\sim$-0.22) correlation between class selectivity and maximum class accuracy drop was observed -- some highly selective units were critical for individual classes, whereas many were unimportant.

These results raise an intriguing question: If selectivity is not predictive of unit importance, what is? This question has been extensively studied in the deep learning literature, primarily with the aim of `pruning' neural networks to generate smaller networks which can run on resource limited devices \citep{lecun1990optimal,Molchanov2016PruningCN}. One of the best signals of feature map importance has been found to be the summed absolute value of weights, suggesting that the magnitude of a feature map may be a deciding factor for importance \citep{Molchanov2016PruningCN}. Pruning remains an active area of research, and a better understanding of the factors which influence unit importance in DNNs may prove to be useful for understanding biological neural networks. 

\section{Dimensionality reduction: How can we characterize distributed representations?} 

If neural networks employ distributed representations \citep{rigotti2013importance,raposo2014category, morcos2016history}, what analysis methods are available for us to make progress towards understanding these representations? A powerful approach which has been applied in neuroscience is to search for low-dimensional structure in large populations of neurons \citep{cunningham2014dimensionality,Gao2017}. The use of low-dimensional codes embedded in high dimensional representations has been hypothesized to be an important organizing principle of neural computation in the brain \citep{GaoGanguli_15,Barrett2016,mastrogiuseppe2017linking}, and so, dimensionality reduction methods are particularly suitable for characterizing neural activity.

Dimensionality reduction methods attempt to identify the neural signals that are maximally correlated with sensory stimuli \citep{archer2014low}, behavioural observations \citep{kobak2016demixed}, neural activity in other brain areas \citep{semedo2014extracting} and instantaneous and temporal covariation within a brain area \citep{YuCunningham_09,MackeBuesing_12,nonnenmacher2017extracting}. Multiple studies have found that neural dynamics and information about stimuli and behaviour are indeed concentrated in low-dimensional subspaces or manifolds \citep{ManteSussillo_13,cunningham2014dimensionality,sadtler2014neural,rabinowitz2015,williams2018unsupervised} and are often distributed across many neurons \citep{machens2010functional,gallego2017neural}. For example, it has been found that dynamics in the motor cortex can be organized into low-dimensional manifolds within a high-dimensional distributed representation \citep{ShenoySahani_13,gallego2017neural}. 

Do these manifolds only reflect correlations in the data, or do they impose constraints on what can be learned? To investigate this question, \citet{sadtler2014neural} used multi-electrode recordings to control a brain-machine interface. By changing the decoder for the brain-machine interface, they demonstrated that animals can learn to adapt to new decoders whenever the required activity patterns remained in the underlying manifold, but could not adapt otherwise. This result suggests that low-dimensional neural manifolds represent functional structure and are not simply a byproduct of statistical correlations. This also has implications for future data-analysis: if the dynamics underlying neural data are low-dimensional and distributed across neurons, a small number of measurements is sufficient for decoding them \citep{GanguliSompolinsky_12,Gao2017}.

Substantial redundancies in representations have also been reported for DNNs. \citet{denil2013predicting} observed that most parameters in DNNs can be predicted from a small number ($\sim5\%$) of parameters without significant reduction in network performance. \citet{li2018measuring} reported that the intrinsic dimensionality of neural networks can be orders of magnitude lower than the number of parameters. Moreover, ablation studies like the ones described in the previous section have  found that more than 85\% of parameters can be removed with minimal performance loss. In addition to linear dimensionality-reduction techniques, nonlinear dimensionality reduction techniques, such as t-SNE \citep{maaten2008visualizing} have also been used extensively to analyze DNNs \citep[e.g.][]{mnih2015human}. 

Relatively little is known about the relationship between DNN performance and the dimensionality of stimulus representations. Recent theoretical work using approaches from statistical physics \citep{chung2018classification} and computational neuroscience \citep{Stringer2018highdimensional} is providing frameworks for understanding the relationship between geometrical properties (including dimensionality) of object-related manifolds and the capacity of deep neural networks to robustly encode information, providing opportunities for development and application of theory-driven data analysis techniques. 

\section{Cross-correlation: How can we compare representations across networks?}

\begin{figure}
    \centering
    \includegraphics[width=.95\textwidth]{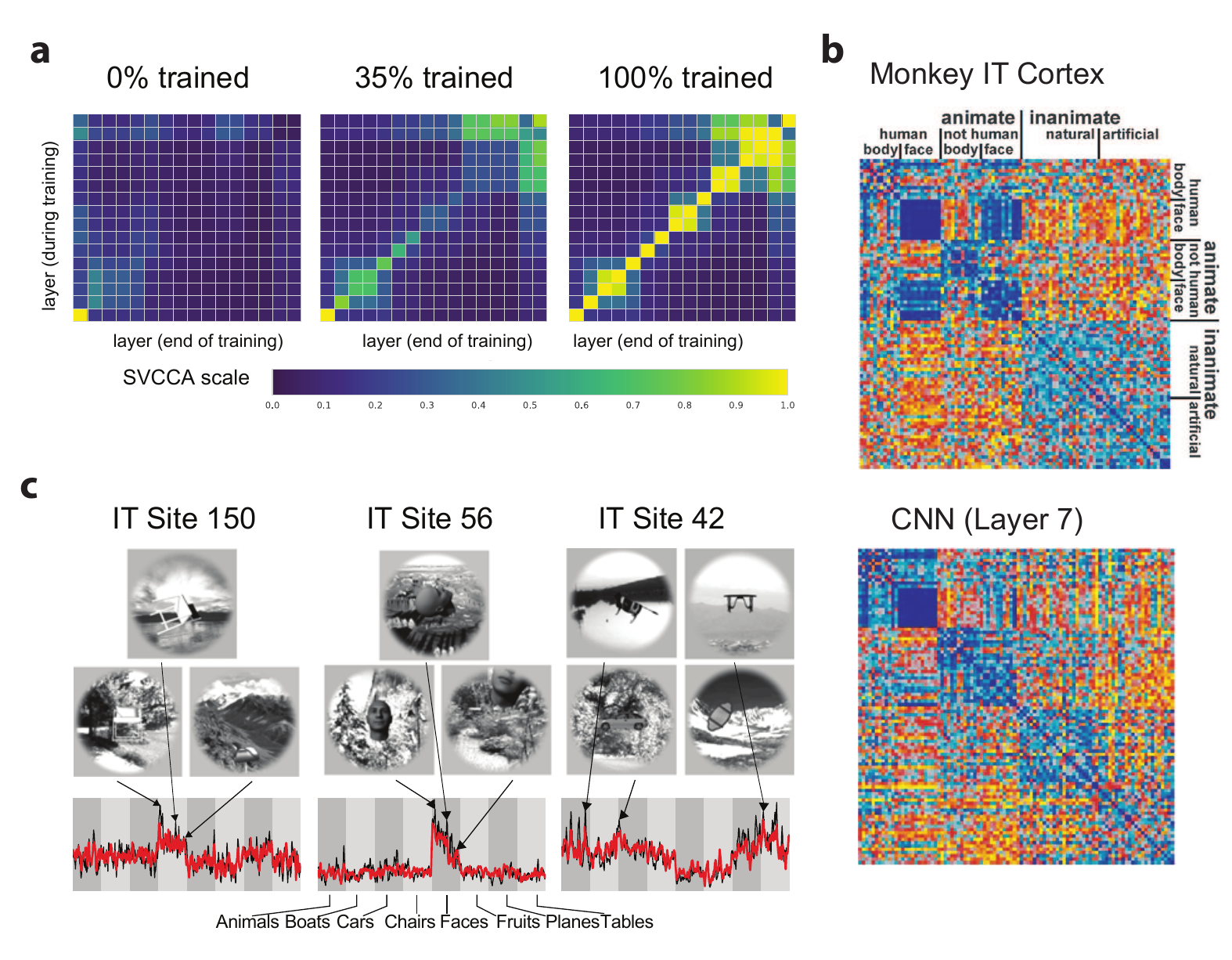}
    \caption{{\bf Statistical methods for comparing representations across networks:} 
    {\bf a)} Learning dynamics in a DNN analyzed using Singular Vector Canonical Correlation Analysis (SVCCA): Each entry in the matrix describes how similar, during training, the representation in each layer is with each other layer after training. One can see that some layers are highly similar to their neighbouring layers, and that some layers converge quickly to their final values. 
    {\bf b)} Representational similarity analysis (RSA) evaluates the (dis)-similarities that two networks assign to different pairs of inputs. The example shows a comparison between recordings in IT and a deep neural network. 
    {\bf c)} Linear regressions can be used to align individual activation vectors between networks. The example shows 3 activation vectors measured in IT cortex (black), and reconstructions from those of a the top layer of a CNN (red).
    \label{fig:comparing}
    Panel {\bf a} was reproduced with modification from \citet{raghu2017svcca}, {\bf b} was reproduced with modification from \citet{khaligh2014deep}, originally published under a CC-BY attribution license: \url{https://creativecommons.org/licenses/by/4.0/legalcode}, and {\bf c} from \citet{yamins2014performance}.
    }

\end{figure}

Just as we may need to go beyond single neuron analyzes toward neural population analysis, we often need to go beyond single population analysis so that we can compare representations across multiple populations of neurons. Such analysis is necessary whenever we need to answer questions about the transformation of representations across layers, or across time. Also, looking towards opportunities for synergy between computational neuroscience and machine learning, it may be useful to compare representations in biological networks directly to representations in artificial networks.

Canonical Correlation Analysis (CCA) is a suitable method for comparing representations across networks \citep{Hotelling1936,semedo2014extracting,raghu2017svcca,Morcos2018cca}. CCA exposes directions which capture correlations between two data-sets (which may contain different numbers of neurons), just as Principal Component Analysis exposes directions in a data-set that maximize variance. \citet{raghu2017svcca} developed an extension of CCA to compare activation-vectors of different neural layers. This analysis revealed low-dimensional structure in distributed representations, consistent with previous dimensionality reduction analysis. Going further, they used this approach to compare different `snapshots' of a network during training and they found that the neural representations in the lower layers of DNNs stabilize much earlier than representations in higher layers (Fig. \ref{fig:comparing} a). Based on this observation, they proposed a new, computationally efficient training-approach in which lower layers are sequentially `frozen' to focus computation-time on higher layers. Finally, they used their CCA-based algorithm to compare two different network architectures trained on the same task (a ResNet and a ConvNet). They found that their representations were similar in early layers, but not in late layers. \citet{li2016convergent} used a related approach to investigate whether DNNs trained on the same task but with a different initial random seed can achieve similar representations. They reported that different networks have units which span overlapping subspaces, even when individual neurons differ. 

Representational similarity analysis \citep{kriegeskorte2008matching}, which  was originally proposed to compare representation between different neural imaging modalities, can also be used to compare networks. In this approach, each network (or layer of a network) is characterized by a `representational similarity matrix' (Fig. \ref{fig:comparing} b), which describes which pairs of stimuli a network considers to be similar or dissimilar.  These pairwise similarities can be calculated from correlation matrices of the corresponding network activations. 

Just as it is possible to compare representations in a DNN with another DNN, it is possible to compare DNN representations directly to biological neural representations. This is particularly interesting, as it opens up the possibility of directly using artificial neural networks as model systems for biological neural networks. This approach was recently adopted by Yamins et al. who used linear regression to identify correspondences between activation vectors of biological neural activity measurements and DNN activity measurements  \citep{yamins2014performance,yamins2016using}. By predicting each `recorded' activation vector from a linear combination of DNN-activation vectors from different layers of a DNN, they found that V4 activity could be best reconstructed from intermediate layers, and IT activity from top layers of the DNN (Fig. \ref{fig:comparing} c). This result is broadly consistent with receptive field analysis in artificial and biological neural networks which indicates that there are similarities between visual processing in a DNN and the ventral stream (Fig. \ref{fig:ReceptiveFields}). However, this requires subjective comparison between receptive fields, unlike cross-correlation based methods, such as CCA, which leverage objective statistical comparisons instead.

\section{Challenges and opportunities}

Neuroscientists and machine-learning researchers face common conceptual challenges in understanding computations in multi-layered networks. Consequently, there is an opportunity for synergy between the disciplines, to redeploy DNN analysis methods to understand biological networks and visa-versa. To what extent is synergy possible, and what challenges need to be overcome?

The first challenge is that DNNs allow full experimental access, whereas this is not possible for biological neural networks. Consequently, many DNN analysis methods can exploit information that is unavailable to neuroscientists. For instance, the activity of all units in a network in response to arbitrary stimuli can be simultaneously measured; the complete weight matrix (``the connectome'') is known; and precise ablation and perturbation experiments can be performed.  Moreover, the full behavioural history of the network (including every stimulus it has ever seen) is known, as is the learning process which determined the weights. Finally, it is usually possible to take gradients with respect to model parameters in DNNs and use these gradients for analysis. 

The second challenge is that, despite both conceptual and empirical similarities \citep{yamins2014performance,khaligh2014deep,kriegeskorte2015deep,gucclu2015deep,yamins2016using} between biological and artificial neural networks at the computational and algorithmic level, there are manifest differences at the mechanistic level. Neural networks used in deep learning are not biologically plausible because, for instance, they rely on the backpropagation algorithm \citep[though see][]{lillicrap2016random,guerguiev2017towards}. Also, they do not produce spike-based representations. The constraints and demands faced by artificial and biological networks are also very different. For instance, brains need to be incredibly power efficient, whereas DNNs must be small enough to fit into computer memory. Whether or not DNNs and biological neural networks use similar representations and algorithms remains an open question. Consequently, analysis methods that may be  informative for DNNs may not be appropriate for biological networks.

A third challenge is that we do not yet fully understand the solutions that DNNs learn, despite having full experimental access in DNNs, and despite them having simpler neural machinery. Since biological networks are substantially more complicated than DNNs, we should expect that it will be even more challenging to understand computation in the brain.

Notwithstanding these challenges, there are ample opportunities for synergy. First of all, several analysis methods for DNNs \textit{can} be applied to biological systems without modification. For example, dimensionality reduction techniques such as CCA only require access to activity recordings.  CCA  could be used to study consistencies in activity patterns in the same neurons over time, across different layers, regions or animals, or to study the similarity of representations across subjects or brain areas. Moreover, a variety of dimensionality reduction algorithms which are specifically suited to biological neural network data have been developed, such as methods that are robust to missing data \citep{balzano2010online}, methods that allow multiple recordings to be combined \citep{turaga2013inferring,nonnenmacher2017extracting}, and methods that are well matched to the statistics of spiking-noise \citep{MackeBuesing_12,gao2016linear} and to  non-stationarities \citep{rabinowitz2015, williams2018unsupervised}.

Second, there is potential for overcoming some limitations that prevent the direct use of DNN analysis methods in neuroscience. For example, for some algorithms which depend on access to gradients, alternative `black-box' variants are being developed which do not require such access and which might enable future application to biological systems, such as recent work on adversarial examples \citep{brendel2017decision}. It will be an important avenue for future work to adapt such methods to the statistical properties of neural activity measurements, and in particular to the fact that typically, we only have access to sparse, noisy and limited, non-stationary measurements in biological neural networks.

Third, DNNs can serve as idealized, in-silico model systems, which can allow researchers to rapidly develop, test and apply new data-analysis techniques on models that can solve  complex sensory processing tasks \citep{chung2018classification,kriegeskorte2015deep}.

Fourth, deep learning is providing new tools for the development of flexible, efficient and powerful neural analysis tools. For example, many Bayesian inference methods were too computationally expensive previously for large-scale data analysis. New approaches for training DNNs to perform Bayesian inference \citep{KingmaWelling_13,PapamakariosMurray_17} are now opening up new avenues to develop efficient Bayesian inference methods for complex, mechanistic models in neuroscience \citep[e.g.][]{gao2016linear,lueckmann2017flexible,speiser2017fast}.

Fifth, an important component of analysis and measurement in machine learning is the use of benchmark data-sets to empirically compare algorithm performance. This approach may also be useful in neuroscience where it can often be difficult to determine, or even quantify, progress. Publicly accessible, large-scale, standardized data-sets are becoming available in neuroscience \citep{hawrylycz2016inferring}, which may enable the development of neuroscience benchmarks and challenges, for example, to predict the response-properties of neurons along the visual hierarchy. These approaches might be useful in comparing, selecting, and ruling out competing models. 

Finally, the fact that it has been difficult to understand DNNs ---  despite full experimental access and the use of simple neural units ---  serves as a reminder that better experimental tools for probing the activity and connectivity of neural circuits are necessary, but not sufficient. Rather, to understand computations in biological neural networks, we additionally require powerful methods for data-analysis, and ultimately we will require quantitative theories explaining these phenomena. Here again, another opportunity for synergy arises between the disciplines. Since DNNs have demonstrated that it is possible for neural systems to support a wide range of complex behaviours, the theoretical insights and understanding that has been developed for DNNs may be directly useful for informing new theoretical neuroscience. For instance, the observation that many of the features of DNNs can arise from very simple underlying principles such as numerical optimization of a  loss-function suggests that many features of biological systems may also be understood from similar underlying optimization principles \citep{marblestone2016toward}. 

The co-incidental revolutions underway in neuroscience and machine learning have opened up a wide array of questions and challenges that have been long beyond our reach. ``Out of adversary comes opportunity'' (B. Franklin), and where there are shared challenges there are opportunities for synergy. At their beginnings, the study of biological and artificial neural networks often confronted these challenges together, and although our disciplines have drifted and diverged in many ways, the time seems to be right, now, to return to this inter-disciplinary collaboration, in theory and in analysis.



\section{Acknowledgements}

We would like to thank Adam Santoro, Neil Rabinowitz, Lars Buesing and Thomas Reber for insightful discussions and comments on the manuscript. We would also like to thank Louise Deason and the DeepMind team for their support. JHM acknowledges support by `SFB 1233 Robust Vision'  of the German Research Foundation (DFG), the German Federal Ministry of Education and Research (BMBF, project `ADMIMEM'), and the Human Frontier Science Program (RGY0076/2018). JHM also thanks his colleagues at research center caesar and TU Munich for  comments on the manuscript. 
\section{Disclosures}

The authors declare no conflict of interest.

\end{document}